# Automatic Reference Models Development: A Framework

Mojeeb Al-Rhman Al-Khiaty*, Moataz Ahmed
Information and Computer Science Department
King Fahd University of Petroleum and Minerals
Dhahran 31261, Saudi Arabia
e-mail: {alkhiaty, moataz}@kfupm.edu.sa

*Abstract*— Software reuse allows the software industry to simultaneously reduce development cost and improve product quality. Reuse of early-stage artifacts has been acknowledged to be more beneficial than reuse of later-stage artifacts. In this regard, early-stage reference models have been considered as good tools to allow reuse across applications within the same domain. However, our literature survey reported in this paper reveals that the problem of automatically developing reference models from given instances has not caught enough researchers' attention yet. Accordingly, in this paper we propose a framework for building a reference model that captures the common and variable analysis/design practices, across the different applications in a domain. The framework considers multi-view models in assessing the commonalities and variabilities among given instances. The proposed framework incorporates learning capabilities to allow improving the quality and reusability of the reference model as it is being used.

## I. INTRODUCTION

The ability to ship a new software product with high quality, within a short timeframe, and with sustainable profit has become vital for software companies to keep up with the new business opportunities [1]. According to Gartner Group Reuse Report [2], software reuse is the only strategy that allows a company to simultaneously address software cost, deliver high quality product, and be competitive in the software market.

*Software reuse* is the process of building new software systems by the use of engineering knowledge or artifacts from existing systems rather than building software systems from scratch [3, 4]. *Systematic software reuse* is the most effective way to significantly improve software development. It reduces the risk of development errors, leverages existing resources, transfers knowledge and experience from experts to the novice, leads to reductions in software development cost and time, and promotes high quality software. Hence, high quality software can be built by incorporating reused artifacts into the software project. Thus, the **goal** of the researchers in software reuse is to come up with systematic procedures for engineering new systems from existing assets [3].

Although the *late-stage* software development artifacts (Code blocks) has been the focus of the software reuse for long time, its recognized benefits have encouraged its practice across the entire software development life-cycle, starting with *domain modeling* through *requirements specification*, *software design*, *coding* and *testing*, to *maintenance* and *operation* [5]. We refer to the first three types of artifacts (*domain modeling*, *requirements specification*, and *software design*) as *early-stage* reusable artifacts while the rest are referred to as *later-stage* reusable artifacts. Reuse at the level of early-stage artifacts has been acknowledged to be more beneficial than reuse of later-stage artifacts [5, 6].

Shifting the engineering focus during system development from late-stage artifacts (i.e. code) to early-stage artifacts (i.e. models) is the aim of *Model-Driven Development* [7]. Model-driven Development (MDD) is a software development methodology which emphasizes the use of models as the primary artifacts in the development process [8]. The level of abstraction, provided by MDD, per se, saves substantial time and resources in production and delivery through: identifying and resolving defects/errors early and thus reducing rework; reusing the design artifacts and knowledge in the subsequent stage (construction) through an automated process [9].

However, during any large-scale development, engineers inevitably have to deal with large collections of models. These models represent different perspectives, different versions across time, different variants in a product family, different applications in a domain, different development concerns and so on [10]. Additionally, these models represent a main source of the knowledge which is captured from the minds of involved people. This knowledge is re-practiced each time new software is created or updated, yet, when comparing software systems, we usually find 60% to 70% of a software product's functionality is common [11]. Thus, without effective reuse mechanism, it is possible to build a new system from scratch, yet a similar situation has been built before. This results in redundant artifacts, and thus redundant maintenance cost and time for the duplicated artifacts. Thus, it is very much needed to have a systematic way to access and reuse existing software models in an efficient way.

One approach with a great potential here is to consolidate these models into a single model. Such a model must be *complete* in the sense that it represents all experience instances (i.e., source models); it must also avoid redundancy by having only one copy of each element that appears in the experience instances. Additionally, the consolidated model must support the desired architectural quality. Moreover, such a model needs to evolve to improve its capability of reuse.

*Software Product Line* (SPL) provides a highly successful paradigm to the thought of consolidation approach. It offers a strategic and promising approach for architecture reuse. Architecture reuse, as a large-grained reuse, has much greater potential than small-grain reuse (e.g., code ruse), as it focuses on reuse of requirements and design [12].

## II. TECHNICAL BACKGROUND AND RESEARCH PROBLEM

SPL architecture is the architecture of a set of a family of products. There are two architectural representations of the product line architecture [12]. The first approach provides a generic architecture for the product line, which captures the commonalities of the products family but ignores all the variabilities. In this approach, each application starts with the generic architecture and adapts it as required.

The second approach, which is more desirable, explicitly captures both the commonalities and variabilities of the products family. Modeling commonalities and variabilities is a key concept in development for reuse. From the reuse perspective, the first architectural representation targets reuse through specialization, as it captures the reusable knowledge and practice at a high level of abstraction [13]. However, this approach fails to capture any knowledge about the variability in a family of products [13]. Moreover, this approach requires a significant effort by experts for specialization [14]. The second architectural representation targets reuse through customization, as it aims at capturing all possible solutions and at the level of details that promotes "as-is" or direct reuse [13]. In this representation, the commonalities among the different possible solutions (artifacts) are unified and represented as common assets and the variabilities are explicitly modeled as alternative (mutually exclusive) or optional assets. Modeling variability in software systems has been acknowledged to be a necessity [15-17]. Variability contributes to the success of reuse in the sense that variable artifacts are modeled to capture the expected diversity in the requirements of the different products while supporting as-is reuse [15].

Therefore, to have a consolidated model (hereafter called *the reference model*) that captures the ruse opportunities of the common and the variable parts across different analysis/design models, the second approach is the one to be adopted from Software Product Line Engineering (SPLE) to the Model Driven Software Development (MDSD).

Synergizing the abstraction capability of MDSD with the variability management capability of SPLE bears the potential benefits of both [18].

However, unless we have enough understanding and experience of the marketing needs about the underlying domain (or similar domain), it is difficult to foresee what is common and what is variable among the product line variants, and thus it becomes difficult to design the software product line upfront [12, 19]. Consequently, developers often create new products in an ad-hoc approach, by using one or more of the available software engineering practices such as copy-and-modify [19]. Following such an ad-hoc approach results in increased unnecessarily duplicated artifacts, increased time-to-market, waste of resources, and losing of opportunities [19]. Additionally, ad-hoc approaches such as the copy-and-modify one may introduce design flaws, or bad smells, into the original models.

Due to the aforementioned issues, proactive product line approach (i.e. SPL first) is rarely used, and usually dominated by reactive (i.e. extending existing SPL) or extractive (i.e. building new SPL from multiple products) approaches. Therefore, when there is a collection of similar software development artifacts the extractive (also called bottom-up) approach is the most applicable to integrate these artifacts in a way that provides an efficient and effective reuse environment [12].

In spite of the considerable efforts that have been made by the researchers towards building a generic artifact out of a set of existing ones, our critical survey revealed that some notable challenges still exist concerning the following building blocks of such consolidation process: 1) the development of an efficient *similarity assessment mechanism* that uses an efficient *comparison algorithm* along with accurate and sufficiently comprehensive *similarity measures* for comparing the different artifacts and identifying their commonalities and variabilities; 2) the development of an efficient *consolidation mechanism* that uses an efficient algorithm for constructing a *generic reference model* that unifies the common artifacts among the different variants and explicates their differences using the similarity information; 3) the development of a mechanism to integrate the *architectural quality* as an orthogonal dimension to the merging process so that the generic reference model consolidation is *guided by* a set of quality factors; 4) a representation mechanism for the *generic reference model* that preserves the necessary information needed for its evolution and its synchronization with the input models; 5) the adoption of a *re-enforcement learning* as a mechanism for continuously improving and readjusting the reference model in a way that makes it a good representative for the input models and meanwhile provides good opportunities for reuse; and 6) providing a tool support to automate the consolidation process.

Addressing the above mentioned challenges is expected to increase the opportunities of early stages reuse, improve the developer productivity, reduce rework, and result in high quality product. Therefore, in this work, we propose a framework for addressing these challenges.

The rest of this paper is organized as follows. Section III provides a technical survey and discussion about generic model-based reuse. Section IV summarizes the main issues related to model consolidation, based on our literature survey. This is followed by the specific research questions in Section V that are meant to be answered by our proposed framework, Section VI.

### III. GENERIC MODEL BASED REUSE

Both SPLE and MDD are emerging technologies that encourage software reuse. The former technology supports the reuse through providing an effective mechanism for reusing the common assets. The later technology (i.e., MDD) supports the reuse through different levels of abstraction provided by the models at different stages of the development life cycle [12, 20]. Adopting the key activities of SPLE into MDD provides a systematic way to build, out of a set of existing MDD models, a reusable reference model with the following benefits [4, 21, 22]:

- Promoting the ruse practice of MDD models from ad hoc into systematic by capitalizing on the commonalities and variabilities managements of an SPLE to capture the commonalities and variabilities across MDD input models.

- Guiding the creation of new applications in a domain.

- MDD models will serve as a reference reusable assets, both horizontally (i.e., for similar products) and vertically (for later stage artifacts).

- The complexity of creating, maintaining, and evolving a set of similar artifacts will be reduced to the simplicity of a single system.

- The model will capitalize on the combined reuse benefits of both SPLE (such as commonalities and variabilities managements) and MDD (such as reducing cognitive distance through model's abstraction).

However, building a reusable reference model out of a set of existing models is not a straightforward task and many issues should be taken into account [10, 23-25]. Among these issues are: detecting the commonality and variability among different models; modeling variability on the merged model, the cohesiveness of the models to be merged; the impact of each input model on the overall quality of the reference model; resolving lexical, structural, behavioral, and semantic conflicts; providing the ability to generate the originating individual models back from the merged model, representing the-state-of-the-art analysis/design practice in the domain, and others.

Different works in the literature have been addressing the problem of consolidating a set of existing models to build a single generic architecture/model. Bayer et al. [26] introduces an approach to support transitioning existing software assets into a product line. Their approach uses a mixture of forward engineering and re-engineering activities.

Bernstein et al. [27] proposes a data model on which the model management functions (matching, selection, merging, and composition) are defined. Their work is an attempt with ultimate objective of establishing a framework for general-purpose model management operators (including matching and merging). However, they highlighted a set of challenges that needs to be tackled towards achieving this objective. Some of these challenges are related to model representation, and the accuracy and the efficiency of both matching and merging operators.

Breivold et al. [1] provided structured migration methods to merge legacy systems to product line architecture based on their industrial experience. In this work they list a set of recommendations for the transition process from legacy systems to the product line. This approach emphasizes the software architecture as a key to recovery of domain concept and relations.

Brunet et al. [28] proposed a framework for research on model merging, in order to be able to discuss and compare the many different approaches to model merging. They propose a set of useful model management operators (*merge, match, diff, split,* and *slice*) and specify the idealized algebraic properties of each operator.

Lutz et al. [24] provide insights into the process of how users compare and merge visual models. The underlying question of their work is "*How do software engineers merge UML models?*" Their main contribution is the use of qualitative theory to demonstrate human model merging activities and the derived findings, as guidelines for tool design. They claim that their findings can be applied to any graph-based, visual models in software engineering. However, the focus in their work is the UML class diagram. They also list a catalog of alternative ways to model the same or similar aspects, in an attempt to show some of the difficulties involved in the similarity assessment and the matching process, which in turns hurts the accuracy of the merge process. The authors also highlight some factors that should be considered when assessing the similarity between models to be merged as well as a set of factors that should be considered by the merging process.

Toward standardizing model merging expectations, Barrett et al. [23] assessed a set of representative merging tools. Their assessment on three merging tools (IBM Rational Software Architect, IBM Rational Rose, Sparx Enterprise Architect) to merge two versions of a simple class diagram showed that the tools "*were not up to the task*" and their performance is "*downright counterintuitive*" even for trivial models. Based on their findings they provide a set of recommendations for the tool vendors. These recommendations are meant to improve conflicts detection and resolution mechanisms, and the accuracy of the merging tool.

Recently, Chechik et al. [10] differentiate three key models' integration operators (merge, weaving, and composition) and describe each operator along with its applicability. Then they elaborate on the merge operator and the factors that one must consider (like, the notation of input models, formalizing the notation, assumptions) in defining a merge operator. They also contrast graph based with semantic based formalization of merge where they argue that graph based approach is often more suitable in earlier stages while semantic based is more suitable in later stages. Additionally, they provide a set of desirable criteria (completeness, non-redundancy, minimality, totality, soundness) to evaluate the merge operator.

The focus of the aforementioned work is mainly on discussing some methodologies, lessons learned, guidelines, challenges, and requirements that should be considered by any comparison or merging algorithm, or tool.

Other work in the literature directed their effort towards proposing and developing different matching and/or merging algorithms and tools [19, 25, 29-34]. Some of these algorithms are specific to particular artifacts [29, 34] and/or specific modeling languages [29] while some others are applicable to more than one type of artifacts [25, 35] and/or more than one type of modeling language [25, 29, 34]. Additionally, these works differ in the information they consider for matching and merging the different artifacts. The following section (Section IV) provides a detailed comparison among these different works.

## IV. MODEL CONSOLIDATION

Model consolidation is the process of merging a set of related models to build a general model that unifies the overlaps of the input models while considering their conflicts and differences. The goal is to provide better model management such as managing evolution [18, 29], [34], managing uncertainty [25, 36], avoiding redundancy, extra cost and/or targeting large-scale reuse [31, 34, 37], migration towards product line from legacy artifacts [19, 20, 26, 33], and views merging [25]. However, as mentioned earlier, building a reusable single model out of a set of existing models involves many issues. In the following we shade the light on the main issues involved in the consolidation process.

### A. Detecting Commonalities and Variabilities

A fundamental operation towards efficient consolidation mechanisms is to have an efficient detection mechanism to identify the commonalities and the variabilities among the models to be merged. There are two main research streams in this area: (1) the development of *similarity measures* (matchers) that adequately capture all the necessary information about the models to be merged; and (2) the development of efficient *matching algorithms* that use the similarity measures to identify identical, similar, and different elements of the models to be merged.

*1) Similarity measures:* there exist a number of similarity measures which can be classified based on the information they capture (universal index [29], name [19, 31, 34, 36, 37], structure [19, 33, 34], layout [31], semantic or role [19, 36], and behavioral [31, 33]), the level of the abstraction (schema-level [30] and instance-level [19, 29, 31, 33, 34, 36, 37]), the level of granularity (element-level [19, 25, 29, 31, 33, 34, 36, 37] and structure-level [34, 36]).

*2) Matching algorithms:* work in this direction can be classified into: tree-based [38], heuristic-based [19], clustering [34] and iterative [29, 34]. Also some matching algorithms can be either exact match [29, 39] or approximate match [19, 31, 33].

### B. Modeling Variants

As mentioned earlier, models are overlapped in some elements while they vary in others. Overlapped elements are unified in the generic model while variants require some mechanisms to track them, understand their touch points and differences, and to be synchronized over time. Work in this direction can fall in two classes: (1) those that model the variants within a single consolidated model, which forms a super set capturing commonalities and variations among the set of input models [19, 25, 29, 31, 33, 34, 36, 37]; and (2) those that keep the variants as separate model fragments [16].

*1) Modeling the variants within a single consolidated model:* in this approach, the consolidated model is characterized by incorporating variation points to distinguish the parts that are common to all variants from those that are specific to certain variants. The idea is to minimize the effort of developing and maintaining model variants by working on a single artifact - the consolidated model- rather than on each variant separately, and then configure the consolidated model via its variation points, so as to obtain one of its input variants when needed. The key issue in this approach is how to represent the variation points. Various approaches exist in literature: (1) using configurable nodes [40]; (2) attaching domain parameters to elements [19, 31, 33]; (3) marking elements with stereo-type or specific notations [25, 36]; (4) using aspect-oriented principles [38, 41]; (5) using feature model notation [19, 37]; (6) using generalization [33, 35]; (7) ordered sequence of changes ($\Delta$) applied to the original model [29, 34], etc.

*2) Modeling variants as separate model fragments:* in this approach variants are modeled as separate model fragments with mechanisms to track their commonalities [16].

### C. Merging Approaches and Algorithms

The goal of any merging algorithm is to combine the input models in such a way that their overlaps are unified to minimize the redundancy among the input models. Merging implies that a comparison of the corresponding elements has been already performed and similarities have been assessed [24]. Work in this direction can be classified into two approaches: (1) Bottom-Up-Top-Down approach [19, 25, 29, 34, 36, 37]; (2) Bottom-Up approach [31, 33, 34].

*1) Bottom-Up-Top-Down merging:* in this approach the merge is performed by the set union of the elements in the input models (Bottom-Up). In other words all the elements in the individual models are presented in the consolidated model. Additionally, it should be possible to generate each one of the input models from the consolidated model (Top-Down). For example, in [19] a Union-merge is used to construct the consolidated model. Additionally, to allow the instantiation of each input model from the consolidated model, a mapping function ($\sigma$) is used to map each element in the input model $M_i$ to its corresponding element in merged model M, and a reverse mappings ($\sigma_1$ and $\sigma_2$) are used to do the reverse (i.e. from M to $M_i$).

*2) Bottom-Up merging:* in this approach the focus is only the merge (Bottom-Up) while replaying the process downward is not considered or guaranteed. For example, in [33] the merged model is refined to become more abstracted using identity and similarity degree threshold. However, no mechanism is provided in the backward direction.

*D. Artifacts and Modeling Language Considered*

Software development involves different artifacts that represent different system perspectives at different level of granularity. The artifacts that are considered by the different approaches are: statechart only [19, 31, 36], class diagram only [29], statechart and class diagram [33], class and sequence diagram [35], feature model [37], goal model and entity relationship diagram [25], process models (activity diagram) [34], etc.

As per the modeling languages for representing the software development artifacts, the matching and merging approaches can be applicable to more than one modeling language [19, 25, 29-31, 34-36] while other approaches are specific to particular modeling language [33, 37]. In the former approaches, models are often represented as generic graphs. This representation makes the match/merge operator generic enough to be applied to different modeling languages. However, these approaches make it difficult to reason about the semantic properties of the merged model. Unlike the generic merge operators, the specific merge operators (often represented as specific graphs) provide a direct basis for reasoning about preservation of semantic properties during merge.

*E. Quality Considerations*

There might be a large collection of models to be merged. Some of these models might be of better quality than the others, though representing the same underlying concept. Therefore, a mechanism is required to detect the different variants that represent the same underlying concept and have a good representation of them in the merged model. Additionally, some mechanism is required to evaluate the quality aspects of the models to be merged. The quality of the software models can be quantified using software metrics [33] or by aligning the models with well-known patterns that are known to support specific quality aspects, e.g., design patterns [42]. The key problem in the first approach is to find the appropriate set of metrics that capture the desired quality [43, 44]. The key problem with the second approach is the detection of such patterns within the input models [45-47].

As we have seen in this section and the previous section, different researchers tried to approach the problem of matching and merging a set of models, with different intentions, considering different types of information, and using verity of algorithms, to build a consolidated model. However, we can highlight an obvious inadequacy with regards to the following issues. The first issue is about the orthogonality of the quality to the merge process. We believe that within the models to be merged there are some patterns that need to be mined. These patterns involve a wealth of knowledge and complex architecture-relevant information that cannot be easily captured by the similarity assessment formulas. Additionally, these patterns target different quality aspects and have specific intents that describe in which situation each pattern should be applied. Therefore, it would be of great value to characterize the input models, through mining the different patterns, and then merge the input models in the light of the quality aspects targeted by those common patterns found within the input models. The second issue is concerning the similarity information. Similarity assessment based on multi-view information (i.e. structural, functional, and behavioral view) is expected to provide more comprehensive and accurate description about the commonality and variability across the different applications within a domain than does the single view information. This makes the reference model more accurate for representing what is common and what is variable among these different applications. The third issue is related to the learnability of the reference model. By learnability we mean the ability of the reference model to learn during its practical use with the objective to improve its quality, usefulness and completeness. It is not unexpected to have a set of experience instances (as source models) that are not fully representatives of the domain for which the reference model is to be built. This will result in a reference model that is not a good representative of the state-of-the-art analysis/design practice of its domain, and thus, hurting the reuse capability of such model. Therefore, a reinforcement learning mechanism is required to improve the reuse capability of the reference model through its practical reuse.

## V. RESEARCH QUESTIONS

The proposed framework is meant to address the following research questions:

1. How can the quality of the input models be improved in the light of a set of well-known analysis/design patterns? What information is needed to identify a pattern or a spoiled version of it?
2. What information is needed to identify commonalities and variabilities across different input models?
3. How can the input models be consolidated to a reference model that represents the input models best meanwhile the desired quality factors are best supported?
4. How should the reference model be represented and annotated so that it preserves the necessary information needed for its evolution and instantiation?
5. How can the reference model learn and evolve when a new experience instances are presented to it or the quality is re-adjusted? How should the learning mechanism be designed to improve the reuse capability of the reference model?

Regarding Research Question 1, some of the source models are expected not to be in a good quality. Thus, a mechanism is required to improve their quality. Our approach to handle this question is explained in Section VI.A. As per Research Question 2, identifying the commonalities and variabilities among the input models is a key to have a reference model that represents all possible solutions (experience instances) with minimum redundancy. Our approach to answer this question (as detailed in Section VI.B) is to use an efficient multi-view-based similarity assessment mechanism that uses an efficient comparison (matching) algorithm (most probably based-on AI techniques, such as Genetic Algorithms) along with a set of multi-view-based assessment measures. Given a collection of design models, the multi-view similarity information between these models, the set of quality factors supported by these models, and the user desirable quality factors, Research Question 3 is concerned with the construction of the reference model so that it represents all possible experience instances with minimum redundancy, less instantiation effort, and it supports the desired quality. The construction of such model requires an efficient merging algorithm that integrates the quality factors along with the similarity information to increase the opportunities of reuse with desired quality supported. Research Question 4 is about the representation and the annotation of the reference model. The commonalities and the variabilities among the different experience instances need to be explicitly represented in the model using a good annotation mechanism. This will also help in both the instantiation and the evolution of the reference model. Our approach for handling these two questions (i.e., Questions 3&4) is explained in Section VI.C. Research Question 5 is concerned with the practical use and the evolution of the reference model. Our approach to answer this question is explained in Section VI.D.

## VI. PROPOSED FRAMEWORK E

Fig. 1 conceptually depicts our proposed framework for a quality-driven reference model (RM) consolidation. The proposed framework can be logically divided into four fundamental components to handle four main phases: preprocessing the source models, assessing the similarity of the input models, building the reference model, and the evolution of the reference model.

### A. Cleaning up the Source Models

It is possible that the input models may have design flaws that can exist in the form of alternatives solutions of the optimal ones, but with degraded quality. These design flaws need to be cleaned up in order to improve the quality of the models to be merged and thus improving the quality of the reference model.

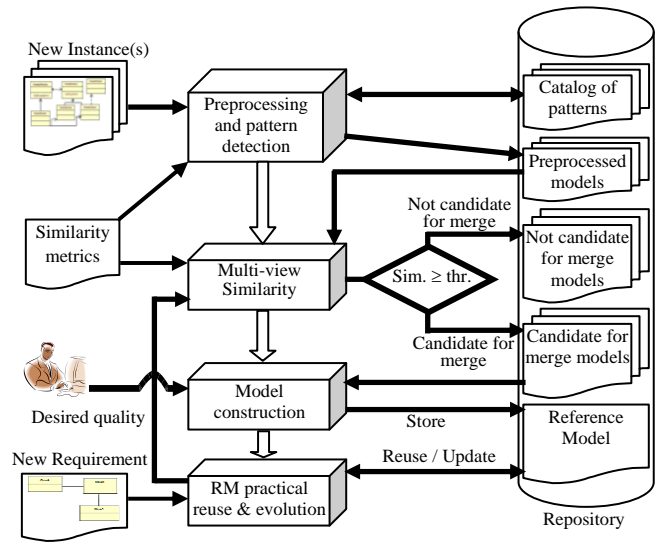

Fig. 1. Proposed framework for quality-driven automatic reference model.

Our approach handles this pre-processing step through aligning these models against a catalog of well-known patterns. Therefore, in our proposed framework, the first step is to detect the different patterns that exist within the input models. The detected patterns might be in its correct form or they might be slightly spoiled (i.e., may have some bad smells and needs to be cleaned). The decision whether a model fragment is a spoiled form of a certain pattern is determined through a similarity assessment between the model fragment and the correct form of the pattern. To make the detection process more efficient we will explore the application of AI techniques to obviate the complexity of the search space. The catalog can be evolved and updated through a re-enforcement learning mechanism.

*B. Similarity Assessment of the Input Models*

After cleaning up the input models (as mentioned in the previous step), the **second** step in our framework is about similarity assessment. In this step, similarities and differences between models are evaluated using an efficient matching algorithm along with a set of accurate and sufficiently comprehensive multiple-view-based similarity measures. Having an accurate and a comprehensive view about each input model, makes the matching of their elements more accurate and thus their commonalities and variabilities are best modeled. Based on this similarity some models may be filtered out, as not candidate for merge, while the rest are passed through to be consolidated in the reference model. The rational behind excluding some input models from being consolidated in the reference model can be justified as follows. Having unrelated models consolidated to a generic reference model, hurts, in addition to other quality aspects, the cohesiveness of the consolidated model. Additionally, following Parnas's thought [48], we can say that the gain of the reference model is of worth when the source models have more elements in common than elements that distinguish them. On the other hand, a large collection of similar or identical models maintained individually also hurts, in addition to other aspects, the maintainability of these models. Additionally, being so restrictive about the models to be consolidated may result in a reference model that may not represent the entire domain. This may hurt its usefulness [35]. Therefore, a balance is required between the cohesiveness of the consolidated model and its generic representation for the input models and the entire domain. Hence, the consolidation process needs to be handled as an optimization problem with the objective of maximizing the reuse opportunities, through maximizing the reusable elements, while keeping the reference model cohesive enough. This requires an efficient selection (based on efficient similarity assessment) of the models candidate for merge. It also requires a feedback mechanism to allow the enhancement of the reference model based on its practical ruse. Our tentative idea for such learning mechanism is detailed in Section D.

*C. Reference Model Construction and Representation*

In this step, the models that are selected as candidate for merge are merged or consolidated to a super-set reference model using an efficient merging algorithm. The commonalities of these models are unified while their diversities are explicated through variation points. Recognizing that the reference model is a reusable asset that captures the recommended practice and knowledge of software systems' artifacts, its quality is of paramount importance. It is clear that targeting the reuse through low quality reference model will offset the expected benefits of reuse and result in low quality products [49]. Therefore, our approach is to develop a consolidation mechanism that uses an efficient heuristic-based algorithm (most probably based on AI techniques) which considers both the multiple-view similarity information and the desired user quality to build such a reference model. The developed mechanism should provide an efficient and effective way for explicitly representing what is common (mandatory) and what is variable (optional or variant) among the different applications. It should also balance the trade-offs between the different desirable criteria, such as, being complete, having minimum redundancy, supporting the desired quality, and being easily instantiatable.

*D. Reference Model Evolution*

After constructing the reference model, it is stored as reusable design artifacts. Bearing the goal of capturing reusable state-of-the-art practices, the reference model needs to be aligned with the new experience instances in the domain to improve its completeness and usefulness for reuse. Additionally, the developer might readjust the quality requirements. These changes require the models (candidates and not candidates) to be re-assessed for similarity and quality, and the reference model to be synchronized and evolve accordingly. Our approach is to handle such evolution and synchronization through developing a re-enforcement learning mechanism that allows us to capture the practical use's feedback of the reference model. For example, when new instance is presented, its similarity to the reference model needs to be evaluated. If this similarity is greater than certain threshold (say, for example, 50%), the new instance can be considered as positive example and it is merged with the reference model, giving the reference model a positive feedback. If, however, the similarity value is less than the threshold value, the corresponding instance is considered as negative example, giving the reference model a negative feedback. In the case of negative feedback, the reference model may need to be evolved to handle the new instance(s). Models that have not been involved in the reference model and marked as "not candidate for merge" need to be revisited during updating the reference model when the update is triggered by the negative feedback. Additionally, some of the instances that are already part of the reference model may show no reuse benefits and/or may hurt the cohesiveness of the reference model and thus need to be excluded from the reference model.

VII. CONCLUSION

In this work we provide a critical survey and discussion about the existing approaches that have been addressing the problem of consolidating a set of existing models to build a single reference model; the information considered to assess the similarity and differences between such models; the requirements that should be considered by the comparison or merging algorithms or tools; and the fundamental issues and challenges involved in such consolidation process. Guided by the findings of our survey, we proposed a framework for building a state-of-the-art reusable reference model that captures the variabilities and commonalities among the different analysis/design instances of different applications in a domain. The proposed framework considers multiple-view similarity information for getting better assessment about the commonalities and the variabilities among the different instances. The proposed framework, also, enriches the proposed reference model with the learning ability to improve its quality and reusability through its practical reuse.

ACKNOWLEDGMENT

The authors would like to acknowledge the support provided by the Deanship of Scientific Research at King Fahd University of Petroleum and Minerals, Saudi Arabia, under Research Grant 11-INF1633-04.

REFERENCES

[1] H. P. Breivold, S. Larsson, and R. Land, "Migrating Industrial Systems towards Software Product Lines: Experiences and Observations through Case Studies," in *Proceedings of Euromicro SEAA'08*, Washington, DC, USA, 2008, pp. 232-239.

[2] Gartner Group, "Software Reuse Report," Stanford1995.

[3] W. Frakes, "Systematic software reuse: a paradigm shift," in *Proceedings of the 3rd International Conference on Software Reuse*, 1994, pp. 2-3.

[4] C. W. Krueger, "Software reuse," *ACM Computing Surveys (CSUR)*, vol. 24, pp. 131–183, 1992.

[5] J. L. Cybulski, "Introduction to Software Reuse," University of Melbourne, Melbourne, Australia1996.

[6] M. Ahmed, "Towards the Development of Integrated Reuse Environments for UML Artifacts," in *The 6th International Conference on Software Engineering Advances (ICSEA'11)*, 2011, pp. 426-431.

[7] B. Selic, "The Pragmatics of Model-driven Development," *Software, IEEE*, vol. 20, pp. 19-25, 2003.

[8] R. France and B. Rumpe, "Model-driven development of complex software: A research roadmap," in *FOSE '07, Future of Software Engineering*, 2007, pp. 37-54.

[9] T. Stahl and M. Volter, *Model-Driven Software Development*: Wiley Publishing, 2006.

[10] M. Chechik, S. Nejati, and M. Sabetzadeh, "A Relationship-Based Approach to Model Integration," *Innovations in Systems and Software Engineering*, vol. 8, pp. 3-18, 2012.

[11] W. Tracz, *Software Reuse: Emerging Technology*. New York: IEEE Press, 1988.

[12] H. Gomaa, Designing Software Product Lines with UML: From Use Cases to Pattern-Based Software Architectures: Addison-Wesley, 2004.

[13] I. Reinhartz-Berger, P. Soffer, and A. Sturm, "A Domain Engineering Approach to Specifying and Applying Reference Models," in *Proceedings of the Workshop Enterprise Modelling and Information Systems Architectures*, 2005, pp. 50-63.

[14] T. Han, S. Purao, and V. C. Storey, "Generating large-scale repositories of reusable artifacts for conceptual design of information systems," *Decision Support Systems*, vol. 45, pp. 665–680, 2008.

[15] M. Sinnema and S. Deelstra, "Classifying variability modeling techniques," *Information and Software Technology*, vol. 49, pp. 717–739, 2007.

[16] S. Apel, F. Janda, S. Trujillo, and C. Kästner, "Model Superimposition in Software Product Lines," in *Proceedings of the 2nd International Conference on Model Transformation (ICMT)*, 2009, pp. 4-19.

[17] L. Chen, M. Babar, and N. Ali, "Variability management in software product lines: a systematic review," in *Proceedings of the 13th International Software Product Line Conference (SPLC '09)*, 2009, pp. 81-90.

[18] A. Pleuss, G. Botterweck, D. Dhungana, A. Polzer, and S. Kowalewski, "Model-driven support for product line evolution on feature level," *Journal of Systems and Software*, vol. 85, pp. 2261–2274, 2012.

[19] J. Rubin and M. Chechik, "Combining Related Products into Product Lines," in Proceedings of the15th international conference on Fundamental Approaches to Software Engineering (FASE'12), 2012, pp. 285-300.

[20] S. D. Kim and H. G. Min, "DREAM: a practical product line engineering using model driven architecture," in *Proceedings of the 3rd International Conference on Information Technology and Applications (ICITA'05)*, 2005, pp. 70-75.

[21] C. W. Krueger and B. Bakal, "Leveraging the Model Driven Development and Software Product Line Engineering Synergy for Success," IBM, White paper 2008.

[22] I. Groher, H. Papajewski, and M. Voelter, "Integrating model-driven development and software product line engineering," in *Eclipse Modeling Symposium*, 2007.

[23] S. Barrett, P. Chalin, and G. Butler, "Model merging falls short of software engineering needs," in *the Proceedings of the 2nd Workshop on Model-Driven Software Evolution, MoDSE '08*, 2008.

[24] R. Lutz, D. Wurfel, and S. Diehl, "How Humans merge UML-Models," in *Proceedings of the International Symposium on Empirical Software Engineering and Measurement*, 2011, pp. 177-186.

[25] M. Sabetzadeh and S. Easterbrook, "View merging in the presence of incompleteness and inconsistency," vol. 11, pp. 174–193, 2006.

[26] J. Bayer, J.-F. Girard, M. Wurthner, J.-M. DeBaud, and M. Apel, "Transitioning Legacy Assets to a Product Line Architecture," in *Proceedings of the 7th ACM SIGSOFT International Symposium on Foundations of Software Engineering (FSE'99)*, 1999, pp. 446–463.

[27] P. Bernstein, H. Halevy, and R. Pottinger, "A vision for management of complex models," *ACM Sigmod Record,* vol. 29, pp. 55-63, 2000.

[28] G. Brunet, M. Chechik, S. Easterbrook, N. Nejati, N. Niu, and M. Sabetzadeh, "A manifesto for model merging," in *Workshop on Global Integrated Model Management (GaMMa'06) co-located with ICSE'06,*, 2006, pp. 5–12.

[29] M. Alanen and I. Porres, "Difference and Union of Models," in *Stevens, J. Whittle, and G. Booch, editors, UML 2003*, 2003, pp. 2-17.

[30] R. Pottinger and P. Bernstein, "Merging models based on given correspondences," in *Proceedings of 29th International Conference on Very Large Data Bases*, 2003, pp. 862–873.

[31] S. Nejati, M. Sabetzadeh, M. Chechik, S. Easterbrook, and P. Zave, "Matching and merging of Statechart specifications," in *Proceedings of the 29th International Conference on Software Engineering (ICSE'07)*, Minneapolis, MN, USA, 2007, pp. 54–64.

[32] K. Bogdanov and N. Walkinshaw, "Computing the structural difference between state-based models.," presented at the Reverse Engineering, 2009. WCRE'09. 16th Working Conference on, 2009.

[33] J. Rubin and M. Chechik, "From Products to Product Lines Using Model Matching and Refactoring," in *Proceeding of SPLC Workshop (MAPLE'10)*, 2010.

[34] C. Li, M. Reichert, and A. Wombacher, "Mining business process variants: Challenges, scenarios, algorithms," *Data & Knowledge Engineering,* vol. 70, pp. 409-434, 2011.

[35] I. Reinhartz-Berger, "Towards automatization of domain modeling," *Data & Knowledge Engineering,* vol. 69, pp. 491–515, 2010.

[36] M. Famelis, R. Salay, and M. Chechik, "Partial models: Towards modeling and reasoning with uncertainty," in *Proceedings of the 34th International Conference on Software Engineering (ICSE)*. 2012, pp. 573-583.

[37] E. A. Aydin, H. Oguztuzun, A. H. Dogru, and A. S. Karatas, "Merging multi-view feature models by local rules," in *the 9th International Conference on Software Engineering Research, Management and Applications (SERA)*, 2011, pp. 140-147

[38] M. Acher, P. Collet, and R. France, "Composing Feature Models," in *the 2nd International Conference on Software Language Engineering (SLE '09)*, 2009, pp. 62–81.

[39] U. Kelter, J. Wehren, and J. Niere, "A Generic Difference Algorithm for UML Models," in *Software Engineering*, Essen, 2005, pp. 105–116.

[40] M. Rosemann and W. M. P. van der Aalst, "A Configurable Reference Modeling Language," *Information Systems,* vol. 23, pp. 1-23, 2007.

[41] M. Voelter and I. Groher, "Product Line Implementation using Aspect-Oriented and Model-Driven Software Development," in *Proceedings of the 11th International Conference on Software Product Line ( SPLC'07)* Washington, DC, USA, 2007, pp. 233–242.

[42] C. Bouhours, H. Leblanc, C. Percebois, and T. Millan, "Detection of Generic Micro-architectures on Models," in *Proceedings of The 2nd International Conferences on Pervasive Patterns and Applications(PATTERNS 2010 )*, 2010, pp. 34-41.

[43] V. Basili and H. Rombach, "The TAME Project: Towards Improvement-Oriented Software Environment," *IEEE Transaction on Software Engineering,* vol. 14, pp. 758-773, 1988.

[44] V. R. Basili, "Software Modeling and Measurement: The Goal Question Metric Paradigm," University of Maryland, Computer Science Technical Report Series 1992.

[45] M. Gupta, R. S. Rao, and A. K. Tripathi, "Design Pattern Detection Using Inexact Graph Matching," in *the International Conference on Communication and Computational Intelligence*, Kongu Engineering College, Perundurai, Erode, T.N.,India, 2010, pp. 211-217.

[46] N. Tsantalis, A. Chatzigeorgiou, G. Stephanides, and S. T. Halkidis, "Design Pattern Detection Using Similarity Scoring," *IEEE Transactions on Software Engineering,* vol. 32, pp. 896-909, 2006.

[47] J. Dong, Y. Sun, and Y. Zhao, "Design pattern detection by template matching," in *Proceedings of the 2008 ACM symposium on Applied computing (SAC '08 )*, 2008, pp. 765-769.

[48] D. Parnas, "Designing Software for Ease of Extension and Contraction," *IEEE Transactions on Software Engineering,* vol. 5, pp. 128–138, 1979.

[49] S. Matook and M. Indulska, "Improving the quality of process reference models: A quality function deployment-based approach," *Decision Support Systems,* vol. 47, pp. 60-71, 2009.

\* Corresponding Author:
Mojeeb Al-Rhman Al-Khiaty,
Faculty of Computer Science and Engineering,
King Fahd University of Petroleum & Minerals,
Dhahran, 31261, Kingdom of Saudi Arabia,
Email: alkhiaty@kfupm.edu.sa    Tel:+966-502079255